\documentclass[conference]{IEEEtran}
\IEEEoverridecommandlockouts
\usepackage{cite}
\usepackage{amsmath,amssymb,amsfonts}
\usepackage{hyperref}
\usepackage{graphicx}
\usepackage{textcomp}
\usepackage{xcolor}
\usepackage{comment}

\usepackage{float}

\usepackage{fancyhdr}
\usepackage{flushend}

\title{Graph Neural Network for Crawling Target Nodes in Social Networks}
\begin{document}



\author{\IEEEauthorblockN{Kirill Lukyanov$^{1,2}$, Mikhail Drobyshevskiy$^{1,2}$, Danil Shaikhelislamov$^{1,2}$, Denis Turdakov$^{1}$}
\IEEEauthorblockA{$^1$\textit{Ivannikov Institute for System Programming of the Russian Academy of Sciences, Moscow, Russia}}
\IEEEauthorblockA{$^2$\textit{Moscow Institute of Physics and Technology (National Research University), Moscow, Russia}}
lukyanov.k@ispras.ru, drobyshevsky@ispras.ru, shaykhelislamov.ds@ispras.ru, turdakov@ispras.ru
}




\pagenumbering{arabic}
\clearpage
\setcounter{page}{31}

\maketitle

\thispagestyle{fancy}
\pagestyle{plain}

\fancyhead{}
\fancyfoot[LO]{\vspace{1em} Published in: 2022 Ivannikov Ispras Open Conference (ISPRAS) \\ DOI: 10.1109/ISPRAS57371.2022.10076873 \textcopyright 2022 IEEE}

\begin{abstract}

Social networks crawling is in the focus of active research the last years.
One of the challenging task is to collect target nodes in an initially unknown graph given a budget of crawling steps.
Predicting a node property based on its partially known neighbourhood is at the heart of a successful crawler.
In this paper we adopt graph neural networks for this purpose and show they are competitive to traditional classifiers and are better for individual cases.
Additionally we suggest a training sample boosting technique, which helps to diversify the training set at early stages of crawling and thus improves the predictor quality.
The experimental study on three types of target set topology indicates GNN based approach has a potential in crawling task, especially in the case of distributed target nodes.

\end{abstract}


\section{Introduction}

Social networks is a rich source of information and they became an object of extensive research over last two decades. However, the access to its data is often severely limited by API or bandwidth restrictions. For instance, Twitter API rate limit for follows lookup are 15 requests per 15 minutes\footnote{\url{https://developer.twitter.com/en/docs/twitter-api/rate-limits}}.
A serious problem when searching for target vertices in real social graphs is the lack of knowledge about the entire graph. In real applications, the problem arises of collecting the maximum number of target vertices with restrictions on the number of queries in an unknown graph.
In our setting, target nodes are all network nodes of interest: influential people, potential employees, or SUV fans. Formally, we suppose there exists an oracle that given a network node can determine whether it is target or not.

\begin{figure}[t]
    \centering
    \includegraphics[width=0.3\textwidth]{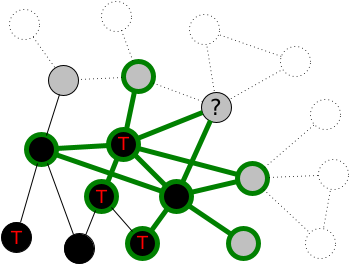}
    \caption{Crawling process. Crawled nodes and edges are black, observed are grey, unknown are drawn dotted. Target nodes have label `T'. Property of the observed node marked with `?' is predicted using its second neighborhood. A GNN predictor structure for this node is indicated with green.}
    \label{fig:crawling}
\end{figure}

In the work~\cite{morales2019deep} a crawling agent is learned to detect tightly connected subgraphs. A deep actor-critic network is pretrained on a synthetic graph with dense target subgraphs and then is applied to a real graph, where it is fine-tuned during crawling.


In our approach, we are also guided by the idea that the property of an unknown node can be predicted from the neighborhood.
The key difference is that we adopt Graph neural network (GNN) models to evaluate the node. Figure~\ref{fig:crawling} illustrates the process for 2-layer GNN. Experiments show that this allows to achieve significantly better results at various types of target sets than using only classical predictors.


In our case we consider three types of target set topology. The first case is one dense subgraph, like is considered in the two mentioned works. The second case is several dense target subgraphs. In practice it can happen that people of interest belong to different communities in a social network.
Finally, target nodes can be uniformly distributed over the social graph without grouping in communities. For example, the distribution of the value of the attribute "gender" on the graph can be considered almost uniform.

In this paper we compare several popular crawlers on the task of collecting a target set. Our main contributions are as follows.
\begin{itemize}

    
    \item We adopt Graph neural networks for a selective harvesting task and show that they allow to crawl more target nodes than classical predictors given the same budget.
    
    \item We develop a sample boosting technique to increase training data available during crawling which improves crawler performance.
\end{itemize}

Source code of our implementation of all the considered methods is publicly available at \url{https://crawling-framework.github.io/}.

In the next section we provide a brief overview of related works. Section \ref{sec:proposed} is devoted to the proposed algorithms and their settings. Then we describe experimental results in section \ref{sec:expr} and, finally, give a conclusion in section \ref{sec:concl}.





\section{Related work}

Areekijseree et al.~\cite{areekijseree2018guidelines} have done a lot of work on the analysis and systematization of various graph traversal methods with respect to the task of collecting as much data (nodes or edges) as possible given a fixed query budget.

Collecting hubs as most influential nodes is a popular special case of selective harvesting in social networks. A series of heuristic approaches were suggested in literature. Avrachenkov et al.~\cite{avrachenkov2014quick} proposed to use the random walk based method to quickly find large degree nodes. They showed experimentally that for large networks the random walk method finds good quality top lists of nodes with high probability. The paper~\cite{shaikhelislamov2020three} proposed a selection algorithm based on the friendship paradox: a random person has fewer friends on average than his friends. Different basic crawlers (RC, MOD, RW, DFS, BFS, DE) are compared in the problem of finding influential vertices~\cite{drobyshevskiy2019collecting}. The results of experiments showed that the leading method for a specific graph may change depending on the query budget. These works showed that the performance of methods strongly depends on the network structure.
Method mostly incorporate heuristics based on structural properties of high-degree nodes rather than learning algorithms, thus they are not applicable to another types of target nodes.


Often, the existing restrictions mentioned in the introduction do not allow to collect the entire graph. Therefore, the majority of graph sampling methods is focused on how to sample preserving graph properties, such as temporal features~\cite{ahmed2013network, ye2010crawling}, scaling properties~\cite{leskovec2006sampling}. These works aimed also at designing sampling schemes specifically to preserve a given quantity. Lee et al.~\cite{lee2006statistical} studied the statistical properties of the sampled scale-free networks, deeply related to the proper identification of various real-world networks. They have investigated the effect of random (non systematic) errors on certain analysis tasks.



A crawler $D^3TS$ solving the selective harvesting task was proposed by Murai et al.~\cite{murai2018selective}. It uses Thompson sampling algorithm over a set of classifiers which is shown to outperform individual classifiers on 5 of 7 real network datasets.

Peter Morales et al.~\cite{morales2019deep} proposed Network Actor Critic (NAC) algorithm. Their approach leverages a Markov decision process formulation of the reinforcement learning to estimate offline models of network discovery strategies (a.k.a. policy) and node utility (a.k.a. reward) that are network state-aware. The model training scheme is as follows. First a detection model is trained on a part of the whole graph and then the model is applied to detect target nodes in the unseen data. As a result, by this inductive learning scheme, the method works well at uncovering dense regions of a network.

A problem related to selective harvesting is known in a web search field as focused crawling~\cite{diligenti2000focused,kumar2021review}. A web crawler chooses the next link to follow based on the information from already visited pages. 
Han et al.~\cite{han2018focused} set a goal to collect as many web pages relevant to the target topic as possible. They use a tabular Q-learning policy that selects the best classifier from a defined pool of classifiers, which evaluate all available candidate URLs that the crawler can immediately fetch. Their MDP formulation somewhat unnaturally allows the agent to deviate from the successive order of states; i.e. it may select actions that do not really exist in a current state. 
Reinforcement learning approaches fits to the sequential nature of the problem, but they cannot be easily adopted for social networks since many features they use are specific to web domain, e.g. incorporate text surrounding a link.

Most of the existing algorithms are adapted for a specific domain and task. Heuristic approaches are also often used, which are also difficult to scale for an arbitrary crawling task. Another serious drawback is that many algorithms use only structural data, which does not allow them to be used to work with attributes. The above reasons lead to the fact that most algorithms in more complex problems begin to behave like Random crawler or Maximal target neighbors.

Another problem is that for $D^3TS$ and  $NAC$ algorithms there is no code or it cannot be run. Therefore, these algorithms are not included in the comparison in the experiments section.

\section{Proposed algorithms}
\label{sec:proposed}

A general scheme of crawling process is the following.
The crawler starts with a single observed node. At each step it uses one predictor to estimate all observed nodes. After crawling this node, the observed nodes and observed graph are updated, predictor is re-trained if needed, and the algorithm updates its weights based on the information whether the new node is target.

Further in this section we describe predictors and feature they use~\ref{ssec:pred}, a sample boosting technique to train predictors~\ref{ssec:boost}.

\subsection{Predictors and features}
\label{ssec:pred}

A predictor is a model that answers how much we are sure a given observed node is target based on the knowledge of the observed graph. Since we want to choose the most promising node among many, we prefer classifiers capable to return a numeric score. Besides, we prefer to use models of different nature in order to overcome the aforementioned tunnel vision effect. We take the following 4 popular models. Gradient boosting classifier (XGB), Random forest (RF), K-nearest neighbors (KNN), and support vector classifier (SVC).
We also consider 2-layers Graph neural networks with 3 kinds of convolutional layers: Graph convolutional network (GCN)~\cite{kipf2016semi}, GraphSAGE (SAGE)~\cite{hamilton2017inductive}, and Graph attention (GAT) with a multi-head attention~\cite{velivckovic2017graph}.

We also add to predictors list a Maximal target neighbors (MTN) heuristic used in~\cite{murai2018selective} as MOD. It chooses the observed node with a maximal number of target nodes among its neighbors.

In this work we consider the first and second neighborhood of the observed node $X$ to predict its property. These neighborhoods comprise all nodes reachable from $X$ in 1 or 2 hops (see Figure~\ref{fig:crawling}). For each node including the observed one, we compute: OD=$1/\sqrt{d}$, where $d$ is its observed degree, clustering coefficient (CC), crawled neighbors fraction (CNF), target neighbors fraction (TNF). For $X$ we additionally compute the fraction of triangles formed with 0, 1, and 2 crawled nodes (Tri0, Tri1, Tri2). Note that all features are in $[0;1]$ interval.
We then compute a 5-bins histogram of these features distribution over first neighbors and normalize it to 1. The feature vector is constructed from OD, CC, CNF, TNF, Tri0, Tri1, Tri2 for the observed node, histograms for OD, CC, CNF, TNF over first neighbors. Triangles for neighbors are not used due to high computation complexity. We also try histograms for second neighbors, but they show no prediction benefits.

Graph neural network is built for every observed node and its two neigborhoods. Each GNN node gets as input a feature vector of 5 elements. OD, CC, CNF for each node and 2 elements encoding whether the node is target ([1, 0] or [0, 1]) except for the observed node which gets [0, 0].

\subsection{Boosting of observed graph for training}
\label{ssec:boost}

Since our task is a sequential decision problem, the predictors are periodically updated using newly obtained information. The main problem is that at early steps there is little information about the graph and target set structure. 
We found it more effective to boost the existing observed graph to obtain more training data.

Suppose the crawler made $n$ steps and got an observed graph $G$. Naturally we have a sequence of $n$ nodes and their neighborhoods obtained during the crawling to train on. Alternatively, the crawler could collect the same set of nodes in different order leading to the same graph $G$, but the sequence of neighborhoods would be different. By simulating alternative possible crawling sequences we increase the training sample size several times. We call this process \textit{sample boosting}.
In order to include neighborhoods of a more relevant size we use for training only around last 20\% of a simulated sequence.

Another point is a possible imbalance of target and non-target nodes in a sequence which can affect predictors quality. To deal with that we exclude excess nodes and add missing nodes from earlier than last 20\% of a sequence starting from the end until the class labels are balanced.

\section{Experiments}
\label{sec:expr}
\subsection{Datasets}

In our experiments we consider three types of graphs different in target set topology (see Table~\ref{tab:datasets}).
\begin{enumerate}
    \item Target set is one dense subgraph.
    \item Target set is several dense subgraphs.
    \item Target set is scattered over the whole graph.
\end{enumerate}

\begin{table}[h!]
\centering
\caption{Datasets used in experiments.}
\begin{tabular}{|l|l|l|l|l|l|}
\hline
\textbf{type} & \textbf{use} & \textbf{graph} & \textbf{target set} & \textbf{budget} & 
\begin{tabular}{@{}c@{}}\textbf{targets} \\ \textbf{fraction}\end{tabular}
\\ \hline
1 & train & donors & top comm. & 100 & 6.1\% \\
1 & train & livejournal & top comm. & 1200 & 0.04\% \\
1 & test & 	kickstarter & top comm. & 700 & 5.6\% \\
1 & test & 	DBLP & top comm. & 1200 & 2.4\% \\
\hline
2 & train & VK-comms & programing & 1000 & 6.5\% \\
2 & train & VK-comms & business & 1000 & 17\% \\
2 & train & VK-comms & health & 3000 & 10\% \\
2 & test & VK-comms & sports & 3000 & 19\% \\
2 & test & VK-comms & philosophy & 3000 & 5.4\% \\
2 & test & VK-comms & citations & 3000 & 9.6\% \\
\hline
3 & train & VK-scatter & sex=1 & 1000 & 50\% \\
3 & test & VK-scatter-L & sex=1 & 3000 & 72\% \\
3 & test & VK-scatter-L & smoking=1 & 3000 & 4.8\% \\
3 & test & VK-scatter-L & smoking=3 & 3000 & 1.4\% \\
3 & test & VK-scatter-L & sex=2 & 3000 & 28\% \\
3 & test & VK-scatter-L & relation=1 & 3000 & 6.8\% \\
3 & test & Twitter & occupation=2 & 3000 & 3.0\% \\
3 & test & Twitter & occupation=7 & 3000 & 0.1\% \\
\hline
\end{tabular}
\label{tab:datasets}
\end{table}

Datasets of type (1) are represented by 4 graphs (donors, livejournal, kickstarter, DBLP) used in~\cite{murai2018selective}, where the target set corresponds to the largest community in a social graph. Sizes of graphs vary from 1.15K nodes in donors to 4M nodes in livejournal.


To deal with the second type we collected our own VK\footnote{\url{https://vk.com}} sample. We chose 6 various topics (business, philosophy, health, etc.), for each topic we randomly picked 5 communities of containing from 100 to 10K subscribers. Then the users and connections between them were collected, excluding private accounts. Finally we extracted the giant graph component containing approximately 100K nodes. 

For the last third type we created another samples of VK network, using a Forest Fire sampling technique starting from a random node. of size 10--100 thousands of nodes. Target sets correspond to all nodes with a certain value of attribute \texttt{sex} (1 or 2), or \texttt{smoking} (takes values from 1 to 7 which encode different relation to smoking).

Some of datasets of each type are taken as training where we develop and fine-tune our algorithms, the rest are used for final experiments.

\subsection{Algorithms and quality metric}
\label{ssec:algs}

For the simulation study we use three types of crawling algorithms.
\begin{enumerate}
    \item Baselines: RC and MTN.
    \item Classical predictors XGB, RF, KNN, SVC,
    \item Graph neural networks: GCN, SAGE, GAT.
\end{enumerate}

Random crawler (RC) chooses a next node randomly from the observed ones. MTN was described in section~\ref{ssec:pred}.

In a predictor mode, the crawler chooses a node with maximal score assessed by the predictor. Corresponding crawlers are named after the predictor they use.


In further experiments we use by default a following standard configuration of parameters unless stated otherwise.
Features: OD, CC, CNF, TNF, Tri for observed node and for first neighbors as a 5 bins histogram.

Predictors: XGB --- default parameters\footnote{Other parameters of XGB, RF, KNN, and SVC are defaults from \texttt{sklearn} library.},
RF --- 100 estimators,
KNN --- `kd\_tree' algorithm, \texttt{n\_neighbors=30},
SVC --- `rbf' kernel,
GCN --- layers sizes (5,5,2),
SAGE --- layers sizes (5,5,2), type of aggregator `gcn',
GAT --- layers sizes (5,5,2), number of heads 3. For GAT, we used a variation with 1 and 3 attention heads, we settled on the variant with 3 attention heads, since it showed more stable results. Further, everywhere GAT means the option with 3 attention heads.
For all the GNNs used\footnote{GNNs are implemented in \texttt{DGL} library~\cite{wang2019deep}, not mentioned parameters are chosen as default.} \texttt{epochs=200}, \texttt{batch=100}, \texttt{learn\_rate=0.01}.

Training strategy: \texttt{train\_from\_size}=10 (before this step predictors are not yet trained),
\texttt{retrain\_step\_exponent}=1.15 (predictors are retrained next time at step $\lfloor s*1.15\rfloor$),
\texttt{train\_max\_samples}=300 (maximal number of training samples obtained by boosting).
Additional boost parameters:
\texttt{max\_boost\_iterations}=20 (maximal number of crawling sequence re-sampling per one training step),
\newline
\texttt{last\_boost\_steps\_fraction}=0.2 (fraction of a re-sampled crawling sequence to be used for training).

For all graphs listed in Table~\ref{tab:datasets} we use a specified budget of steps.
The crawler starts with the knowledge of one target node. This is done to avoid possible long random walking before it finds the first target node.
When the budget is exhausted we measure the number of target nodes collected by the crawler. After performing multiple runs starting from a random target node, we compute the median. In contrast to the mean, the median is insensitive to outliers occurring when the starting node is separated from the other targets.

\subsection{Feature selection}

Before testing crawlers with multiple predictors, we search for optimal configurations of individual predictors and fine-tune boosting parameters.

First, we compare several combinations of features used with classical predictors XGB, RF, KNN, and SVC:
\begin{enumerate}
   \item TNF;
   \item OD,CC,CNF,TNF;
   \item OD,CC,CNF,TNF for observed and average over first neighbors;
   \item Tri (here and further means including all 3 of Tri0, Tri1, and Tri2);
   \item OD,CC,CNF,TNF,Tri;
   \item OD,CC,CNF,TNF,Tri for observed and average over first neighbors;
   \item OD,CC,CNF,TNF,Tri for observed and 5 bins histogram over first neighbors;
   \item OD,CC,CNF,TNF,Tri for observed and 5 bins histograms over first and second neighbors.
\end{enumerate}

Due to a limited space, we give only the implications.

\begin{itemize}
    \item Adding OD,CC,CNF to TNF improves quality (1-10\% more target nodes crawled).
    \item Only 3 Tri features give a good quality.
    \item Adding OD,CC,CNF,TNF to Tri can improve quality but not significantly (not more than 2\%).
    \item Adding OD,CC,CNF,TNF to the second neighbors does not significantly affect.
    \item Using 5 bins histogram instead averaging can slightly improve quality in average.
\end{itemize}

As a compromise we choose combination (7) of features and use it by default.

As for GNNs, we also varied their feature set. We conducted experiments where appended 1-hot encoded VK profiles attributes (age, occupation, marital status, etc) to the node input vector, but this did not increase the number of crawled nodes.

\subsection{Tuning boosting parameters}

Here we determine the influence of two boosting parameters, fraction of last steps taken for training and amount of training data in comparison to historical training.
Both historical and boosting strategies have 2 more parameters that could be fine tuned, when to start training and how often to retrain. We choose the former to be 10, since at earlier steps the observed graph is too small, and the graph indicators are statistically incorrect. The optimal frequency of retraining is exponential with exponent 1.15. According to the preliminary experiments, retraining more often or even at each step does not give a notable profit but takes much more computational time. This becomes critical at high budgets while the size of the observed graph grows super-linearly.

First we vary the size of training samples generated by boosting. We take \texttt{train\_max\_samples} equal to 300, 1000, 3000, and compare to the historical regime. We run a crawler with XGB, RF, KNN, SVC predictor on training graphs of 3 types. Results are averaged over 20 runs. Since all graphs are quite different, we provide here only the aggregated results.
To determine the best parameter we range the results for each graph and each predictor. The leader is given score 1, the second place got score 0.5. Then we sum the scores over graphs, see Table~\ref{tab:boost_size}. All boosting configurations have higher total scores than the historical version, \texttt{train\_max\_samples}=300 is the best one. Interestingly, merely increasing training sample via boosting does not provide a higher predictor quality as one could expect.

\begin{table*}[h!]
\centering
\caption{Performance of crawlers with predictor (MTN, XGB, RF, KNN, SVC) and GNNs crawlers (GCN, GAT, SAGE) on test graphs. Best 2 algorithms for each graph are in bold.}
\begin{tabular}{l|ccccccccccc}

Graph  & \textbf{MTN} & \textbf{XGB} & \textbf{RF} & \textbf{KNN} & \textbf{SVC} & \textbf{SAGE} & \textbf{GCN} & \textbf{GAT} \\ \hline
donors & \textbf{36$\pm$1} & 33$\pm$6 & 33$\pm$5 & \textbf{35$\pm$3} & 31$\pm$6 & 30$\pm$2 & 30$\pm$2 & 34$\pm$1 \\
livejournal & 602$\pm$6 & 683$\pm$23 & \textbf{707$\pm$11} & 680$\pm$12 & 666$\pm$153 & \textbf{752$\pm$8} & 658$\pm$55 & 529$\pm$154 \\
dblp & 585$\pm$19 & 638$\pm$84 & \textbf{656$\pm$56} & 579$\pm$36 & 650$\pm$35 & \textbf{901$\pm$13} & 610$\pm$52 & 388$\pm$48 \\
kickstarter & 158$\pm$72 & 155$\pm$21 & 231$\pm$16 & \textbf{260$\pm$14} & \textbf{249$\pm$16} & 220$\pm$20 & 151$\pm$57 & 138$\pm$90 \\ \hline
VK-prog & 386$\pm$88 & 509$\pm$162 & 603$\pm$192 & 532$\pm$139 & 583$\pm$221 & \textbf{737$\pm$34} & 672$\pm$72 & \textbf{777$\pm$65} \\
VK-busi & \textbf{913$\pm$76} & 867$\pm$94 & 904$\pm$19 & 872$\pm$19 & \textbf{927$\pm$85} & 785$\pm$21 & 681$\pm$57 & 642$\pm$76 \\
VK-heal & 2597$\pm$97 & 2699$\pm$35 & 2701$\pm$98 & \textbf{2702$\pm$66} & \textbf{2737$\pm$41} & 2471$\pm$20 & 2458$\pm$216 & 1933$\pm$108 \\
VK-phil & 1391$\pm$32 & 1746$\pm$286 & 1834$\pm$87 & 1741$\pm$204 & 1818$\pm$118 & \textbf{1960$\pm$13} & \textbf{1892$\pm$164} & 1847$\pm$140 \\
VK-sport & 2311$\pm$422 & 2533$\pm$277 &\textbf{ 2866$\pm$124} & 2752$\pm$374 & \textbf{2827$\pm$272} & 2465$\pm$12 & 2047$\pm$116 & 2481$\pm$222 \\
VK-cita & \textbf{2207$\pm$153} & 1829$\pm$206 & 1623$\pm$252 & 1620$\pm$435 & 1911$\pm$231 & \textbf{2043$\pm$48} & 1344$\pm$97 & 1928$\pm$189 \\
\hline
VK-sex & 569$\pm$142 & 757$\pm$32 & 783$\pm$28 & 696$\pm$28 & \textbf{811$\pm$44} & \textbf{840$\pm$17} & 635$\pm$105 & 739$\pm$116 \\
VK-sex1 & 2157$\pm$145 & 2366$\pm$50 & 2539$\pm$17 & 2318$\pm$14 & \textbf{2644$\pm$45} & \textbf{2540$\pm$47} & 2200$\pm$49 & 2357$\pm$235 \\
VK-smok1 & 199$\pm$41 & 150$\pm$9 & 182$\pm$3 & 169$\pm$12 & 203$\pm$9 & 267$\pm$2 & \textbf{306$\pm$5} & \textbf{340$\pm$4} \\
VK-smok3 & 74$\pm$14 & 51$\pm$8 & 50$\pm$7 & 54$\pm$9 & 59$\pm$11 & \textbf{102$\pm$3} & 45$\pm$4 & \textbf{83$\pm$9} \\
VK-sex2 & 1501$\pm$141 & 1479$\pm$54 & 1043$\pm$21 & 1397$\pm$18 & 1252$\pm$41 & \textbf{2003$\pm$43} & 1120$\pm$60 & \textbf{2081$\pm$225} \\
VK-relat1 & 148$\pm$10 & 240$\pm$14 & 199$\pm$9 & 196$\pm$18 & 234$\pm$21 & \textbf{273$\pm$7} & 270$\pm$25 & \textbf{278$\pm$13} \\
TW-occup2 & \textbf{787$\pm$317} & 324$\pm$154 & 226$\pm$85 & \textbf{794$\pm$325} & 251$\pm$76 & 275$\pm$34 & 224$\pm$41 & 440$\pm$315 \\
TW-occup7 & 13$\pm$3 & 21$\pm$6 & 11$\pm$2 & 19$\pm$5 & 12$\pm$1 & 10$\pm$0 & \textbf{32$\pm$6} & \textbf{43$\pm$8} \\
\end{tabular}
\label{tab:all}
\end{table*}

\begin{table}[h!]
\centering
\caption{Varying of max number of training samples generated by boosting. Each cell is a total score over 5 training graphs: leader gets 1, second place gets 0.5. The rightmost column is the sum in the row.}
\begin{tabular}{l|cccc|c}
\textbf{} & \textbf{XGB} & \textbf{RF} & \textbf{KNN} & \textbf{SVC} & \textbf{total} \\ \hline
historical &  0  &  2  &1.5&  0  &  3.5 \\ \hline
300        &  3  & 4.5 & 3 &  4  & \textbf{14.5} \\ \hline
1000       &  2  & 0.5 & 2 &  2  &  6.5 \\ \hline
3000       & 2.5 & 0.5 & 1 & 1.5 &  5.5 \\ 
\end{tabular}
\label{tab:boost_size}
\end{table}

Next, we compare different values for the fraction of last boosting steps used for training. We take \texttt{last\_boost\_steps\_fraction} as 0.2, 0.5, 0.8, and 1.0 and also compare them with the historical version. Using the same aggregation scheme, we find optimal \texttt{last\_boost\_steps\_fraction} to be 0.2, see Table~\ref{tab:boost_frac}. This parameter does not show monotonic dependency, which could probably be explained by highly stochastic individual results.

\begin{table}[h!]
\centering
\caption{Varying of fraction of last boosting steps used for training. Each cell is a total score over 5 training graphs: leader gets 1, second place gets 0.5. The rightmost column is the sum in the row.}
\begin{tabular}{l|cccc|c}
\textbf{} & \textbf{XGB} & \textbf{RF} & \textbf{KNN} & \textbf{SVC} & \textbf{total} \\ \hline
historical &  0 &  1 &  2 &  1  &   4 \\ \hline
0.2        &  2 & 2.5&  3 &  3  & \textbf{10.5} \\ \hline
0.5        & 2.5&  1 & 1.5& 0.5 &  5.5 \\ \hline
0.8        & 2.5&  2 &  1 &  2  &  7.5 \\ \hline
1.0        & 0.5&  1 &  0 &  1  &  2.5 \\ 
\end{tabular}
\label{tab:boost_frac}
\end{table}

\subsection{Comparison of all algorithms}

Now we compare algorithms (description in section\ref{ssec:algs}) at the same task. 3 GNN models are: GCN, SAGE, and GAT, the last has 3 heads in multi-head attention. Layers sizes are the same for all GNNs and are equal to 5 (input dimension), 5 (hidden layer), and 2 (output classes).

For XGB, RF, KNN, and SVC, the standard set of features is used, while GNNs input features are OD, CC, CNF, and 2 elements encoding whether the node is target (except for the observed one).

\begin{figure}[h!]
    \centering
    \includegraphics[width=0.5\textwidth]{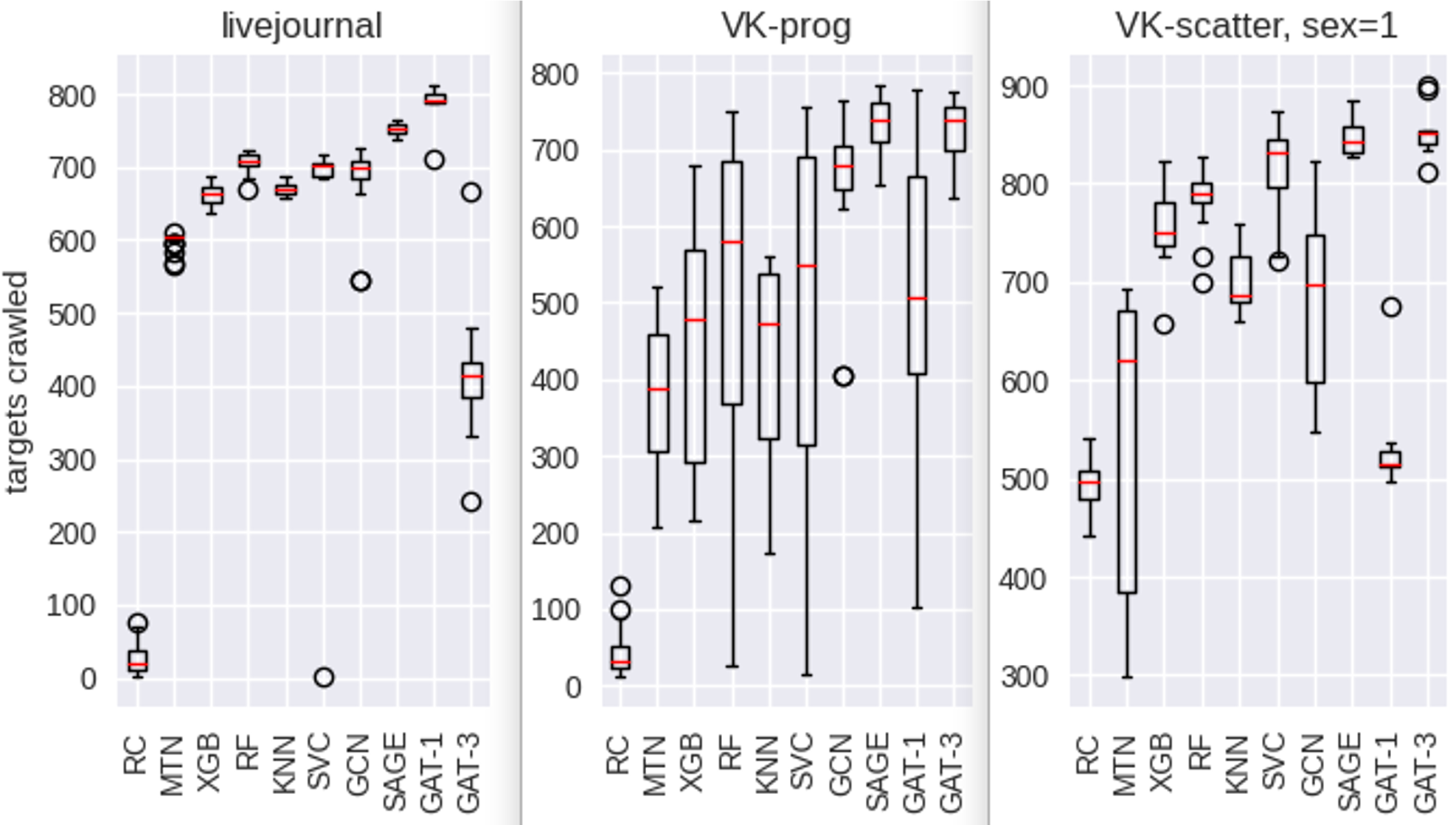}
    \caption{Comparison of crawlers on graphs of each type. GNNs achieve better results than classical predictors and have less variance.}
    \label{fig:features}
\end{figure}

Figure~\ref{fig:features} shows how many nodes is crawled on training graphs of each type. One can see that some GNNs predictors achieve equal or higher results than all classical ones. Moreover, at VK-prog all classical algorithms show very high variance, while GNNs have much more stable results.

For a final comparison on test graphs we take a baseline MTN, 4 crawlers with predictors (XGB, RF, KNN, SVC) and 4 crawlers with GNN (GCN, SAGE, GAT). 

Table~\ref{tab:all} gives the results. KNN, SVC and GAT crawlers are ahead the others on 6 graphs from 22, SAGE --- on 10. One can see that there are no better crawlers for the selected graph types.

Another observation in first type of graphs SAGE is the leader twice. In second type of graphs, SVC and SAGE often shows a better result than the other algorithms considered. On the last group of graphs, SAGE and GAT show the best results. In general, SAGE showed good results on all types of problems, and GAT is competitive with another leader - SVC.

\section{Conclusion}
\label{sec:concl}

In this paper we have presented two novel contributions.
The first one is sample boosting for better training of models in the course of the crawling algorithm.
Secondly, we have shown the effect of applying graph neural networks in the task of crawling target nodes.
It also looks promising to investigate more complex configurations of GNNs, since at the moment there are suggested a plenty of different types of layers for graph neural networks.

A perspective future research direction is to apply several predictors guided by an online learning method. We believe it could intelligently switch between the classifiers during the crawling choosing the most appropriate predictor to use at the moment.

\bibliography{reference}
\bibliographystyle{IEEEtran}

\end{document}